\documentclass[default,iicol]{sn-jnl}

\usepackage{natbib}
\usepackage{graphicx}
\usepackage{subfig}
\usepackage{mathbbol}
\usepackage{amssymb}
\usepackage{eqnarray,amsmath}
\usepackage{bm}
\usepackage{pifont}
\usepackage{wasysym}
\usepackage[symbol]{footmisc}
\usepackage{xparse}
\usepackage[nameinlink,noabbrev]{cleveref}

\theoremstyle{thmstyleone}%
\theoremstyle{thmstyletwo}%

\DeclareMathOperator{\Tr}{Tr}
\NewDocumentCommand\dsum{e{_^}}{{\displaystyle\sum_{#1}^{#2}}}

\DeclareSymbolFontAlphabet{\amsmathbb}{AMSb}%

\begin{document}

\title[\quad]{Shear-thickening of dense bidispersed suspensions}


\author*[1]{\fnm{Alessandro} \sur{Monti}}\email{alessandro.monti@oist.jp}
\equalcont{These authors contributed equally to this work.}

\author*[1]{\fnm{Marco Edoardo} \sur{Rosti}}\email{marco.rosti@oist.jp}
\equalcont{These authors contributed equally to this work.}

\affil*[1]{\orgdiv{Complex Fluids and Flows Unit}, \orgname{Okinawa Institute of Science and Technology}, \orgaddress{\street{1919-1 Tancha}, \city{Onna, Kunigami District}, \postcode{904-0495}, \state{Okinawa}, \country{Japan}}}


\abstract{We study the rheological behaviour of a dense bidispersed suspension varying 
the relative size of the two dispersed phases. The main outcome of our 
analysis is that an enhanced flowability (reduced relative viscosity) of the 
suspension can be achieved by increasing the dispersion ratio of the phases. 
We explain the observed result by showing that the presence of large 
particles increases the packing efficiency of the suspension, leading to a 
reduction of the contribution of the contacts on the overall viscosity of 
the suspension in the shear-thickening regime, i.e. where the contacts are 
the dominating component.}


\maketitle

\section{Introduction}
Dense suspensions of rigid, spherical particles immersed in a Newtonian fluid tend 
to exhibit a lower flowability, acting almost like a rigid solid when subjected 
to an increasing shear stress. This behaviour, which is well-known in the framework 
of non-Newtonian fluids, undergoes the name of shear-thickening 
\citep{FREUNDLICH1938,BARNES1989,MELROSE1996,WAGNER2009,MEWIS2012}. 
Shear-thickening suspensions are becoming increasingly important in industry, 
with preferential applications in shock absorbers, such as soft-body armours, 
or into smart fluids with rheological properties that can be tuned by acting 
on the mechanisms of the stress transmission \citep{LIN2016,CLAVAUD2017,OZTURK2020}. 
The growing relevance of such fluids has stimulated the recent progress made 
to unravel the mechanisms that dominate the thickening behaviour. In particular, 
simulations and experiments have shown that, when an increasing stress that 
exceeds a critical value is applied to a dense suspension, the rhelogical 
response due to the interacting particles switches from mainly lubricative 
to chains of frictionally contacting grains that transmit the stress and 
reduce the flowability of the suspension 
\citep{SETO2013,FERNANDEZ2013,WYART2014,GUY2015,LIN2015}. Shear thickening has 
been classified into two regimes, named after the macroscopic behaviour of 
the dense suspensions: continuous shear thickening (CST), i.e. identifiable 
as a smooth increase of the viscosity with the shear rate, and discontinuous 
shear thickening (DST),  i.e. an abrupt transition to a solid-like suspension 
with reduced flowability 
\citep{BROWN2009,BROWN2011,SETO2013,FERNANDEZ2013,MARI2015}. 
The two behaviours have been observed for several kinds of dense suspensions 
that span from monodisperse to polydisperse in terms of size of the beads 
\citep{CWALINA2016,MADRAKI2018,RATHEE2021}, and made of Brownian 
\citep{BARNES1989,MARI2015} or non-Brownian particles 
\citep{LERNER2012,WYART2014,MARI2014,MONTI2021}.

Considering bidispersed suspensions, 
several parameters can be used to control
the shear-thickening regime. Surely, one of those is the friction coefficient
of the particles, that can be tuned to increase or reduce the viscosity 
in the thickened regime \citep{HSU2018,JAMALI2019,MORE2020,PRADEEP2021}. 
Among the other parameters,
the dispersion ratio and the relative 
percentage of volume fraction of the two dispersed phases can also be 
used to modify the shear-thickening behaviour
\citep{BENDER1996,JAMALI2013,MADRAKI2017,MADRAKI2018}.
However, within the literature, there are interpretation of experimental 
results that look in contrast with the theory of maximally random jammed (MRJ) 
packings \citep{TORQUATO2000,DONEV2004} and of the densest binary sphere 
packings \citep{HOPKINS2012,HOPKINS2013}.
The latter show that adding particles of different sizes is 
a more efficient way of packing particles, with an increased MRJ packing, 
thus suggesting, in the case of flowing suspensions, an increased flowability 
and a postponement of the onset (critical value of the stress) of the DST when 
gradually enhancing the total volume fraction of the suspension. This was also
observed by \citet{DHAENE1994}, who characterized the rheological
behaviour of bimodal colloidal suspensions.
On the other side, recent experimental studies \citep{MADRAKI2017,MADRAKI2018}
show that, by adding an increasing volume fraction of large 
particles, the suspension can be successfully led to a transition from CST to 
DST, thus resulting in a larger suspension viscosity at a fixed shear rate
in the presence of large particles.  The authors proposed a geometrical 
explanation based on an excluded-volume effect to justify the experimental 
observations: the presence of large particles causes the formation of shells 
around them (with size linked to the radius of the small ones), mostly composed 
of aqueous solution since the small beads are hindered of densely packing 
around a large sphere, resulting in a local decrease of the concentration 
of the small particles within the volume of the shells. This local decrease 
of volume fraction is compensated by an increased concentration in other 
regions of the suspension, due to mass-conservation, that triggers the 
discontinuous thickening \citep{MADRAKI2017,MADRAKI2018}. 

\begin{figure*}[t]
\includegraphics[width=\textwidth]{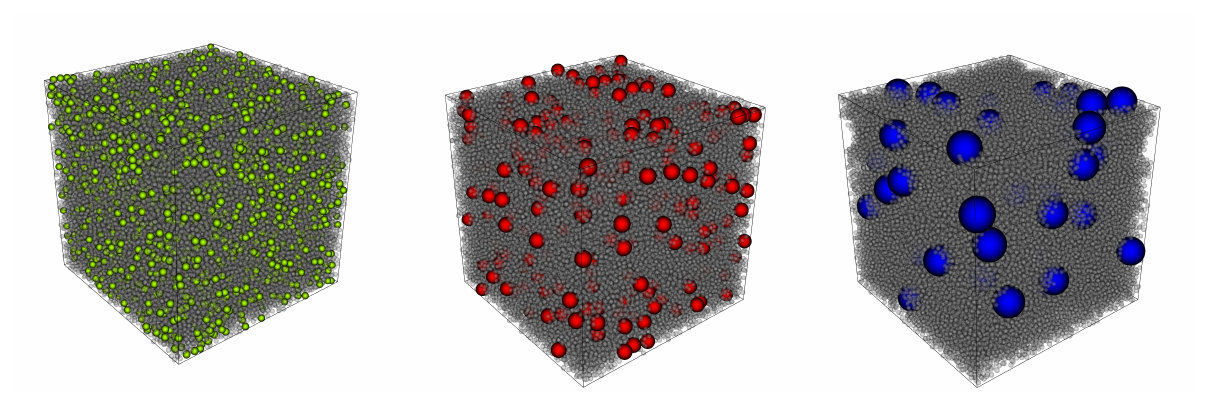}%
\caption{\label{fig:susp} Visualization of bidispersed dense suspensions at 
    volume fraction $\phi=0.50$, for different size of the larger particles 
    dispersed (increasing from left to right). The grey colour shows the smaller 
    particles, while the green, red and blue colours show the larger particles 
    for the suspension with dispersion ratio $\lambda=\{1.5,3.0,6.0\}$, 
    respectively. This colour-scheme will be used for referring to the different 
    configurations throughout the whole manuscript.}
\end{figure*}


In this work, we reconcile the apparent contradiction between the literature data
on the subject by showing that the suspension viscosity 
reduces when mixing large particles in a suspensions at a fixed total volume 
fraction and shear rate.  As a consequence, we show that a dense bidispersed 
suspension with large dispersion size exhibits a milder shear thickening regime 
compared to a suspension having identical volume fraction and a lower 
dispersion ratio, consistent with the higher efficiency of the MRJ packing 
shown by the highly bidispersed suspensions \citep{HOPKINS2013}. 
Studies on the effect of the particles sizes were carried out in 
the past \citep{CHANG1993,CHANG1994a,CHANG1994b,PEDNEKAR2018}. 
These works numerically tackled the effect of bidispersion by varying 
the dispersion size
and the relative volume ratio between the large and small dispersed
particles. A decrease in viscosity has been recorded as the suspension moved
from monodisperse to bidispersed, with a minimum taking place when the 
$\phi_2/\phi=0.65$, where $\phi_2$ is the volume fraction of the 
large dispersed phase and $\phi$ is the volume fraction of the suspension
\citep{PEDNEKAR2018}. A further decrease in viscosity has been observed when
the dispersion ratio increases, i.e. increasing the size of the larger dispersed
phase compared to the smaller one \citep{PEDNEKAR2018}. The suspensions 
studied in those works, however, contained a relatively small number of 
particles $n \le 2000$, and no variation of the shear-rate has been
included. Here, we extend the previous studies by increasing the number 
of particles of the suspensions and by introducing the shear-rate
dependency, specifically studying the effect of bidispersion on the
shear-thickening.

\section{Methodology}
The problem at hand has been tackled by numerically investigating the effects on 
the shear-thickening region for three configurations of bidispersed suspensions of 
frictional particles subjected to a uniform shear-flow, with dispersion size 
$\lambda=a_2/a_1$ in the range $\{1.5, 3.0, 6.0\}$ ($a_1$ and $a_2$ being the 
radius of the small and large suspended phase, respectively). 
The total number of particles is set to $n=2^{16}$, with the number of particles 
of the two dispersed phases determined by fixing the relative volume between 
them equal to $V_2/V_1=0.25$; four total volume fractions of 
the suspension are 
considered: $\phi=\{0.45,0.50,0.55,0.60\}$. 
To extend the validity of the results, for the suspension with $\phi=0.60$, 
we considered three relative volumes of the dispersed phases, such that 
$\phi_2/\phi=\{0.2, 0.5, 0.8\}$, where $\phi_2$ is the volume fraction of the larger 
phase.
\Cref{fig:susp} shows a representation of the three dispersion sizes considered, 
with the larger phase highlighted with a colour-scheme that will be kept through 
the whole manuscript.

\begin{figure*}[htb]
\centering
\includegraphics[width=\textwidth]{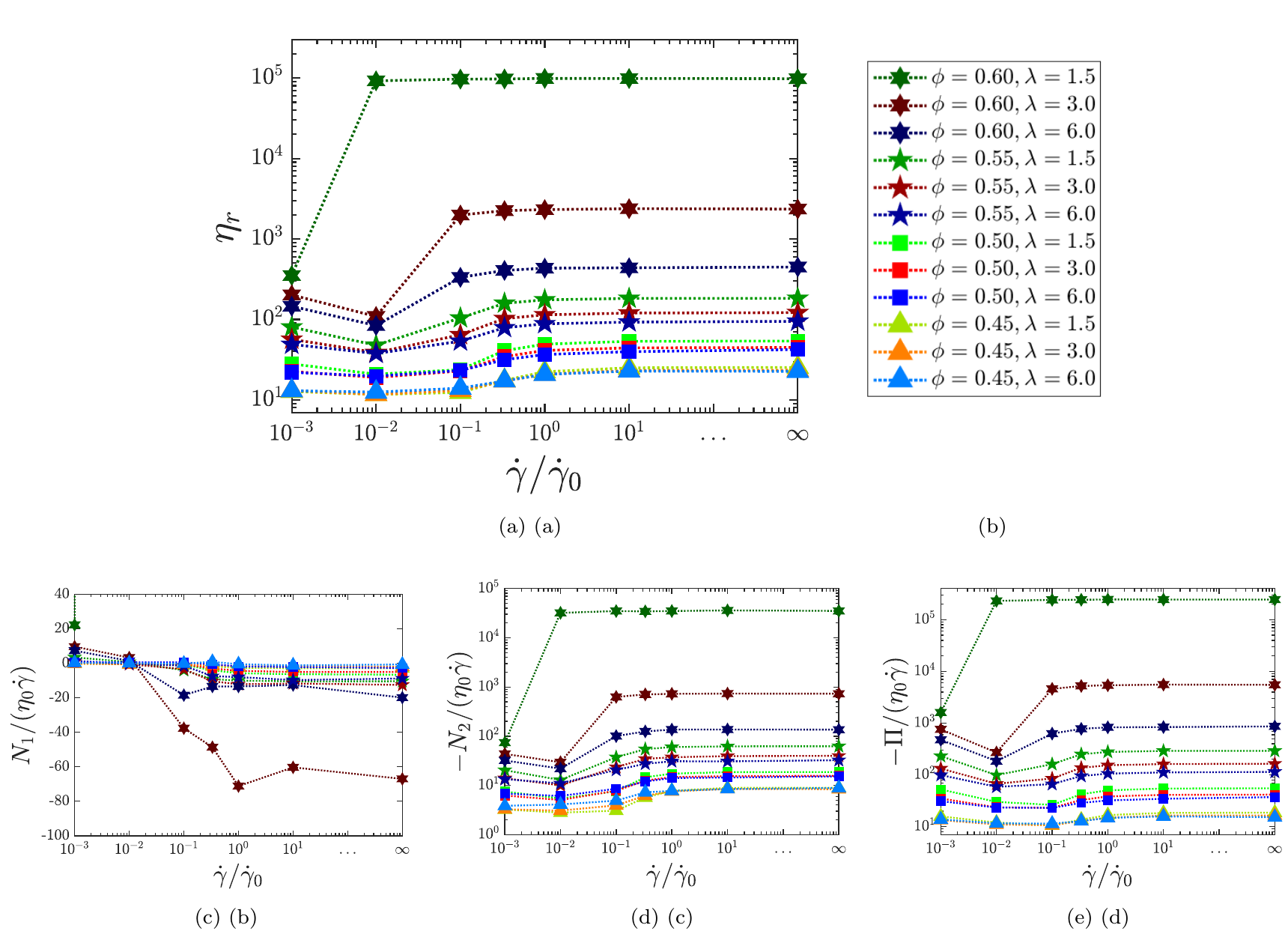}%
\caption{\label{fig:etaRgd} Panel (a): Relative viscosity $\eta_r$ as a function of 
    the non-dimensional shear-rate $\dot{\gamma}/\dot{\gamma}_0$. 
    Panel (b): First normal stress difference $N_1/(\eta_0 \dot{\gamma})$ as 
    a function of the non-dimensional shear-rate $\dot{\gamma}/\dot{\gamma}_0$.
    The curve for the case ($\phi=0.60$, $\lambda=1.5$) largely exceeds
    the limits of the figure,  reaching positive values of about $N_1\sim 8000$,
    and are not shown to maintain the other curves visible.
    Panel (c): Second normal stress difference $-N_2/(\eta_0 \dot{\gamma})$ as 
    a function of the non-dimensional shear-rate $\dot{\gamma}/\dot{\gamma}_0$.
    Panel (d): Pressure $-\Pi/(\eta_0 \dot{\gamma})$ as a function of    
    the non-dimensional shear-rate $\dot{\gamma}/\dot{\gamma}_0$.
    In all the figures, the curves are obtained for suspensions at volume fraction 
    $\phi=\{0.45,0.50,0.55,0.60\}$, denoted by the triangle, square, star 
    and star hexagon markers, respectively. The shades of green, red and blue represent 
    the different dispersion ratio $\lambda=\{1.5,3.0,6.0\}$, as shown 
    in \cref{fig:susp}. The markers and the colours are summarized within
    the legend shown in the top right of the figure.}
\end{figure*}
The numerical investigation is carried out using a validated and publicly 
available software, \textit{CFF-Ball-0x} \citep{MONTI2021}. 
\textit{CFF-Ball-0x} is an in-house software that models a dense, non-Brownian 
suspension (volume fraction $\phi\ge 0.40$) of quasi-inertialess, neutrally-buoyant, rigid 
spherical particles in a shear-flow defined by the shear-rate $\dot{\gamma}$.
The code tackles the Newton-Euler equations that govern the translational 
and rotational dynamics of the rigid particles,
\begin{equation}
    \begin{cases}
        m_i \dfrac{\mathrm{d}\boldsymbol{u}_i}{\mathrm{d}t} &=
    \dsum_{j=1}^{N_H} \boldsymbol{F}^H_{ij} +
    \dsum_{j=1}^{N_C} \boldsymbol{F}^C_{ij} +
    \dsum_{j=1}^{N_E} \boldsymbol{F}^E_{ij}, \\
    \mathbb{I}_i \dfrac{\mathrm{d}\boldsymbol{\omega}_i}{\mathrm{d}t} + 
        \boldsymbol{\omega}_i \times (\mathbb{I}_i \boldsymbol{\omega}_i) &=
    \dsum_{j=1}^{N_H} \boldsymbol{T}^H_{ij} +
    \dsum_{j=1}^{N_C} \boldsymbol{T}^C_{ij} +
    \dsum_{j=1}^{N_E} \boldsymbol{T}^E_{ij}, \\
    \end{cases}
    \label{eq:NE}
\end{equation}
where the subscript $i$ indicates the particle $i\in [1,N]$, being
$N$ the number of particles. The right-hand side of System \eqref{eq:NE}
lists the forces and torques resulting from particle-flow and 
particle-particle interactions with the $j$th neighbour, and 
the superscripts $H$, $C$ and $E$ indicate the nature of contribution,
i.e. hydrodynamics, inelastic contacts and electro-chemical effects, 
respectively. The forces and torques are applied to the centre of mass 
of the $i$th particle, with mass $m_i$ and inertia tensor 
$\mathbb{I}_i$ and cause a variation in the translational 
and angular velocities of the particles, here denoted by the symbols 
$\boldsymbol{u}_i$ and $\boldsymbol{\omega}_i$, respectively.

Next, we describe the models implemented for the three contributions
listed above. Concerning the hydrodynamics of the system, 
dense suspensions of rigid particles immersed in a low-Reynolds-number 
flow experience a Stokes drag and a pair-wise, 
short-range lubrication force \citep{MARI2014}, caused by the 
relative motion of near particles that squeezes the fluid flowing 
in the narrow gaps between them. The two contributions are implemented
as a linear relationship between forces (torques) and velocities
(angular velocities) \citep{BALL1997,MARI2014}, 
$\boldsymbol{F}^H=-\mathbb{R}(\boldsymbol{u}-\boldsymbol{U}^\infty)$,
where $\mathbb{R}$ is a resistance matrix that is obtained by
neglecting the far-field effects and considering only the dominant 
near-field divergent elements coming from the squeeze, shear and pump 
modes, following the work by Mari \textit{et al.} \citep{MARI2014}.

The contact contribution is modelled with the stick-and-slide 
method \citep{LUDING2008}.
This method mimics the inelastic contacts with 
spring-dashpot systems that start operating when two spheres overlap.
The force can be expressed as $\boldsymbol{F}^C=(k_n\delta + 
\gamma_n\dot{\delta})\boldsymbol{n} + k_t\xi \boldsymbol{t}$;
the spring-dashpot systems are oriented along the normal (centre-to-centre)
$\boldsymbol{n}$ and tangential $\boldsymbol{t}$ directions (the latter 
to model the frictional component of the contacts) and are 
characterised by the spring and dashpot constants, 
($k_n$,$\gamma_n$) and ($k_t$,$\gamma_t=0$), respectively, where
the subscript indicate the direction ($n$ normal and $t$ tangential).
The values $\delta$ and $\xi$ indicate the normal and tangential
displacements of the springs and $\dot{\delta}$ the normal projection
of the overlapping velocity. The tangential contribution is subject 
to the Coulomb's law, $\lvert \boldsymbol{F}^C_t\rvert \le \mu 
\lvert \boldsymbol{F}^C_n\rvert$, where $\mu$ is the frictional coefficient.
In this work, the friction coefficient is set to $\mu=0.5$ 
for all the scenarios analysed, as we did in a previous work where we compared
our results with experimental data \citep{RATHEE2021}.

\begin{figure*}[htb]
\centering
\includegraphics[width=\textwidth]{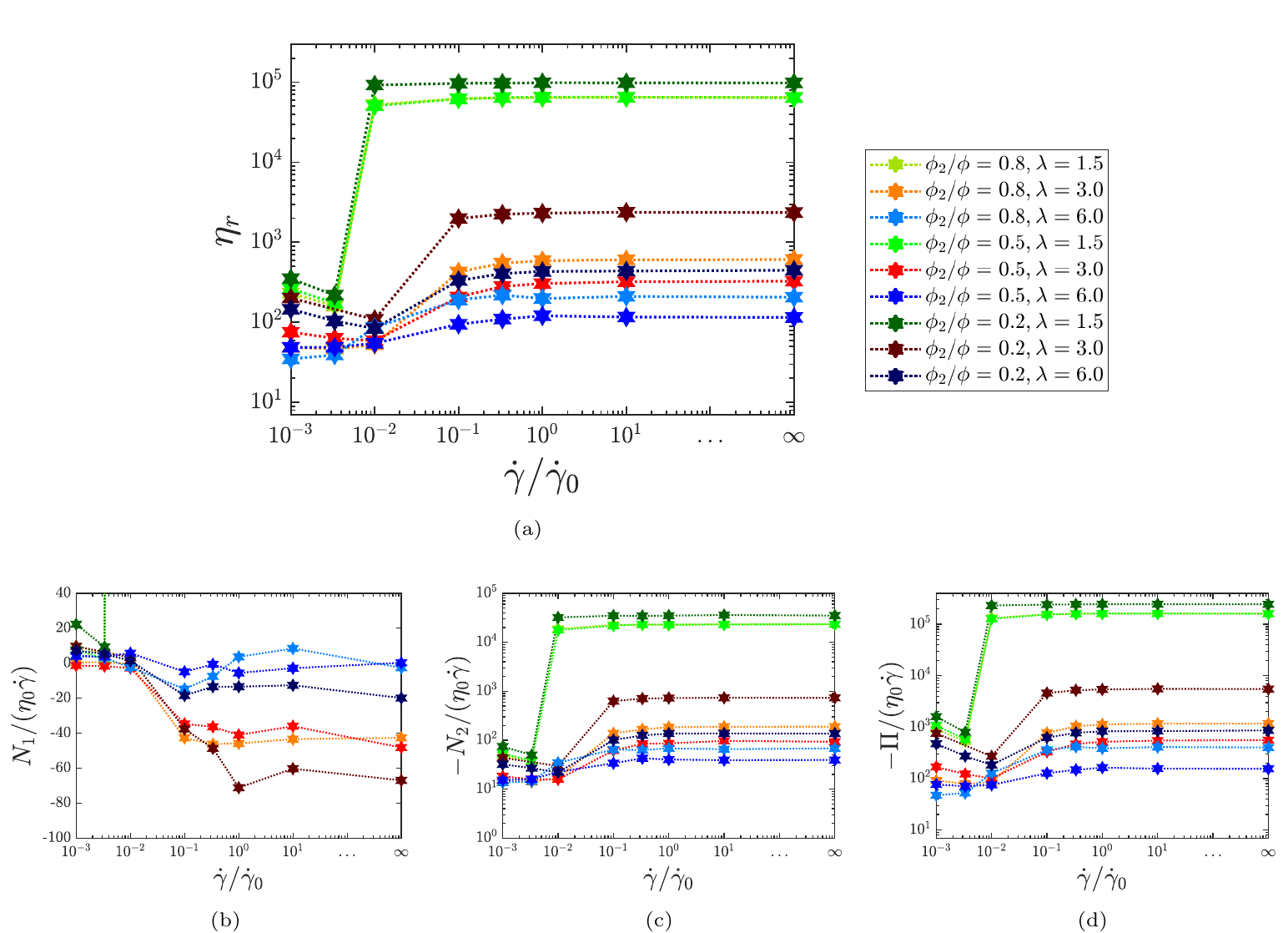}%
\caption{\label{fig:etaRgdRev} Panel (a): 
    Relative viscosity $\eta_r$ as a function of 
    the non-dimensional shear-rate $\dot{\gamma}/\dot{\gamma}_0$. 
    Panel (b): First normal stress difference $N_1/(\eta_0 \dot{\gamma})$ as 
    a function of the non-dimensional shear-rate $\dot{\gamma}/\dot{\gamma}_0$.
    The curve for the case ($\phi=0.60$, $\lambda=1.5$) largely exceeds
    the limits of the figure,  reaching positive values of about $N_1\sim 8000$,
    and are not shown to maintain the other curves visible.
    Panel (c): Second normal stress difference $-N_2/(\eta_0 \dot{\gamma})$ as 
    a function of the non-dimensional shear-rate $\dot{\gamma}/\dot{\gamma}_0$.
    Panel (d): Pressure $-\Pi/(\eta_0 \dot{\gamma})$ as a function of    
    the non-dimensional shear-rate $\dot{\gamma}/\dot{\gamma}_0$.
    In all the figures, the curves are obtained for suspensions at volume fraction 
    $\phi=0.60$ (hexagon markers) varying the relative volume fractions 
    of the dispersed phases, with $\phi_2/\phi=\{0.2, 0.5, 0.8\}$, 
    from darker colours to lighter ones.
    The shades of green, red and blue represent 
    the different dispersion ratio $\lambda=\{1.5,3.0,6.0\}$, as shown 
    in \cref{fig:susp}. The markers and the colours are summarized within
    the legend shown in the top right of the figure.}
\end{figure*}
The last contribution considered in this work are forces 
that arise from the electrochemical properties of the suspension.
These contributions are modelled as a sum of an
inter-particle, distance-decaying repulsion force, 
and an attraction force in van der Waals form
\citep{GALVEZ2017,SINGH2019}, 
$\boldsymbol{F}^E = \boldsymbol{F}^R+\boldsymbol{F}^A$, where the
superscripts $E,R,A$ stand for electrochemical, repulsive and 
attractive, respectively. The force resulting from the repulsive 
contribution can be written as 
$\boldsymbol{F}^R=F_0 e^{-h/L_s}\boldsymbol{n}$, where $F_0$ is
the magnitude of the force, $L_s$ is the screening length and $h$
is the particle-particle surface distance, while the attractive
force can be written as $\boldsymbol{F}^A = 
A\overline{a}/12(h^2+\varepsilon^2) \boldsymbol{n}$, where $A$ is
the Hamaker constant, $\overline{a}$ is the harmonic mean radius of 
the two particles involved and $\varepsilon$ is a smoothing term 
to eliminate the singularity when the two particles touch, i.e.\ 
$h=0$.

Several parameters dominate the physics of the suspension described by
the system of equations \eqref{eq:NE}. Considering the translational equation
of \eqref{eq:NE} and gathering the dimensional quantities, we can
apply the Buckingham's $\Pi$ theorem and find the set of the
non-dimensional groups that can be tuned to outline the desired
dynamics of the suspension. In particular, by choosing as
independent fundamental quantities the scales that define the
hydrodynamics, the three following non-dimensional groups emerge:
the Stokes number 
\begin{equation}
    St = \rho_p a_0^2 \dot{\gamma}/\eta_0,
\end{equation}
that arises from the inertial term and compares
the time-scale of the particles with the time-scale of the hydrodynamics.
In the equation above, $\rho_p$ is the density of the particles (equal to
the density of the carrier fluid in the cases we considered), $a_0$
is the typical length scale of the system (the radius of the particles of
the smallest phase of the suspension), $\eta_0$ is the viscosity of 
the carrier fluid and $\dot{\gamma}$ is the shear-rate applied to the 
suspension. The constraint $St \ll 1$ is applied to enforce the 
inertialess regime. The second number is the non-dimensional 
stiffness
\begin{equation}
    \hat{k}=k_n/(\eta_0 a_0 \dot{\gamma}),
\end{equation}
that weighs the importance of the contacts contributions compared 
to the hydrodynamic term. In this case, the constraint $\hat{k}\gg 1$ 
is set to force the particles to 
be rigid. We considered the normal spring constant $k_n$ as the 
dominant term of the spring-dashpot systems since we enforce the 
additional constraints $k_t=2/7k_n$ and $\gamma_n\dot{\gamma}/k_n
\ll 1 \sim O(10^{-7})$, where the latter is the relaxation time of 
the spring-dashpot system. Finally, the last non-dimensional group,
i.e.\ the equivalent shear-rate
\begin{equation}
	\hat{\dot{\gamma}} = \lvert \boldsymbol{F}^E (d_{ij}=0)\rvert /(\eta_0 a_0^2 \dot{\gamma}) =
    \dot{\gamma}_0/\dot{\gamma},
\end{equation}
refers to the time-scale introduced by the electrochemical 
contribution and is tuned to set a shear-rate dependent rheology 
on the suspension. Here $d_{ij}$ is the relative surface-surface distance
between the particles $i$ and $j$.

More details can be found in our previous publication \citet{MONTI2021}, 
together with validations against theory, numerical simulations 
and experimental measurements. 
The software can be freely downloaded at the following link
https://github.com/marco-rosti/CFF-Ball-0x.\\
\begin{figure*}[htb!]%
\includegraphics[width=\textwidth]{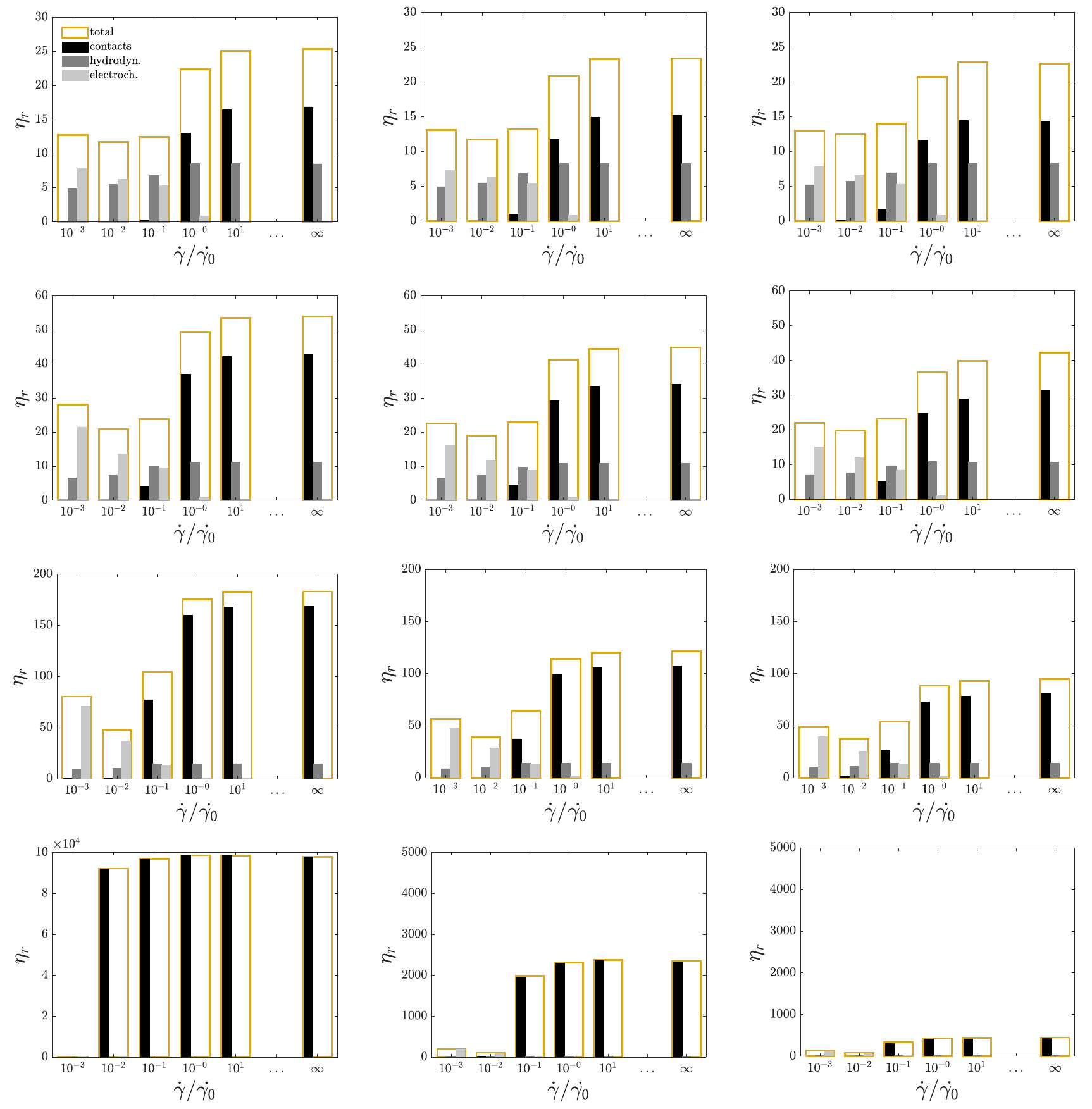}%
\caption{\label{fig:cont} Contributions to the relative viscosity $\eta_r$ 
    of the contacts (black), hydrodynamics (darker grey) and electrochemical 
    potentials (lighter grey), as a function of the non-dimensional 
    shear-rate $\dot{\gamma}/\dot{\gamma}_0$. The suspensions represented 
    have increasing volume fractions from the top row ($\phi=0.45$) to the 
    bottom row ($\phi=0.60$) and increasing dispersion ratio, 
    $\lambda=\{1.5,3.0,6.0\}$, from left to right. Note that, in the last row,
    the bars in the left panel are on a different scale compared with the
    bars in the central and right panels.}
\end{figure*}%

\section{Results}

We start the analysis by looking at the rheological properties of the suspensions.
Panel (a) of \cref{fig:etaRgd} shows the flow curve of the suspension, 
i.e.\ how the relative viscosity $\eta_r$ of the suspension varies with the 
applied shear rate $\dot{\gamma}/\dot{\gamma}_0$.
The curves are grouped using the markers to 
distinguish the different volume fractions analyzed (triangles: $\phi=0.45$; 
squares: $\phi=0.50$; stars: $\phi=0.55$; star hexagons: $\phi=0.60$) 
and the colours to discriminate 
the dispersion ratio, in accordance with \cref{fig:susp} (green: $\lambda=1.5$; 
red: $\lambda=3.0$; blue: $\lambda=6.0$). This is recapped in the legend of
\cref{fig:etaRgd}. All the curves plotted show 
a shear-thinning region followed by the 
a Newtonian plateau at low shear-rates, followed by a region of 
viscosity increase, i.e.\ the shear-thickening regime, and a final plateau 
at high shear rates. We observe that the presence of large particles reduces, 
at high shear-rates, the relative viscosity of the suspension and weakens 
its shear-thickening behaviour: thus, we find that a bidispersed suspension 
with large dispersion ratio exhibit CST where an equivalent suspension (with 
same volume fraction) of monodispersed particles would have an abrupt change 
of the rheological properties, e.g.\ DST. The effect of the dispersion ratio 
is enhanced for large volume fractions, as getting closer to the jamming.
These findings are consistent with those by \citet{DHAENE1994},  but different
from those by \citet{MADRAKI2017,MADRAKI2018}. A possible explanation of this
discrepancy has to be traced back to the increase of the total volume fraction 
of the suspension in the latter \citep{MADRAKI2017,MADRAKI2018}.

When studying the rheology of a suspension, the two independent
normal stress differences must be considered, along with the viscosity. 
The values obtained by the numerical simulations of these quantities together 
with the pressure 
are shown in \cref{fig:etaRgd}, panels (b)--(d), where panel (b) shows the first 
normal stress difference $N_1 = \Sigma_{11}-\Sigma_{33}$, 
panel (c) the second normal stress difference $N_2 = \Sigma_{22}-
\Sigma_{33}$, and panel (d) the pressure 
$\Pi = \Tr{(\Sigma)}/3$; in the definition of the rheological quantities, 
$\Sigma$ represents the stress-tensor computed by means of the stresslet
theory \citep{GUAZZELLI2011}. The three rheological quantities are shown 
for all the cases studied in this work.
Overall,  we observe that the first normal stress difference is small and dominated by fluctuations, 
as already observed by other authors \citep{MARI2014,SINGH2019}. 
The second normal stress difference is always negative and its absolute value
exhibit a trend similar to the pressure and relative viscosity, 
consistently to the ones observed in the literature \citep{MARI2014}.
The effect of the dispersion ratio for these two quantities, closely 
reflects the behaviour shown by the relative viscosity, with the exception of the first normal stress 
difference which suddenly changes sign and strongly increases by several orders of magnitude
when the suspension undergoes DST.

After that, we picked the most interesting case in terms of rheological 
properties, i.e. $\phi=0.60$ where $\lambda$ affects DST ($\lambda=1.5$) by 
weakening it to CST ($\lambda=\{3.0, 6.0\}$), 
and we analysed the trends of the relative viscosity,
normal stress differences and pressure varying the relative volume fraction
of the dispersed phase, i.e. $\phi_2/\phi=\{0.2, 0.5, 0.8\}$.
As per \cref{fig:etaRgd}, panel (a) of \cref{fig:etaRgdRev} shows the flow curve 
of the suspension, 
i.e.\ how the relative viscosity $\eta_r$ of the suspension varies with the 
applied shear rate $\dot{\gamma}/\dot{\gamma}_0$.
The curves are grouped using the colours to discriminate 
the dispersion ratio, in accordance with \cref{fig:susp} (green: $\lambda=1.5$; 
red: $\lambda=3.0$; blue: $\lambda=6.0$) and the shades of the colours to
identify the relative volume fractions analysed, with colours going from
light to dark for increasing $\phi_2/\phi$. This is recapped in the legend of
\cref{fig:etaRgdRev} (note that the curves $\phi_2/\phi=\{0.5, 0.8\}$ at $\lambda=1.5$ almost overlap). The curves show, as in \cref{fig:etaRgd}, 
a shear-thinning region followed by the 
Newtonian plateau at low shear-rates, followed by a region of 
viscosity increase, i.e.\ the shear-thickening regime, and a final plateau 
at high shear rates. The effect of the relative volume fractions controls the 
relative viscosity, reducing it when the two phases have similar volume fractions
and increasing it when the dispersion tends towards the monodispersion. For the cases
with $\phi_2/\phi=\{0.2, 0.8\}$, consistently
with the previous works \citep{CHANG1993,CHANG1994a,CHANG1994b,PEDNEKAR2018,GUY2020}, the
relative viscosity is higher when there are more smaller particles than larger ones
immersed in the suspension.
Even in this case, we observe that the relative volume fraction can reduce, 
at high shear-rates, the relative viscosity of the suspension and weakens 
its shear-thickening behaviour, especially when $\phi_2=\phi_1$. However, the 
effect of the relative volume fraction does not have a strong impact as the 
the relative size of the dispersed phases.

The normal stresses of these suspensions, instead,  
are shown in \cref{fig:etaRgdRev}, panels (b)--(d), where panel (b) shows the first
normal stress difference $N_1 = \Sigma_{11}-\Sigma_{33}$,
panel (c) the second normal stress difference $N_2 = \Sigma_{22}-
\Sigma_{33}$, and panel (d) the pressure
$\Pi = \Tr{(\Sigma)}/3$, as in \cref{fig:etaRgd}; 
in the definition of the rheological quantities,
$\Sigma$ represents the stress-tensor computed by means of the stresslet
theory \citep{GUAZZELLI2011}. The three rheological quantities are shown
for all the pairs of relative volume fractions $( \phi_1, \phi_2 )$  
at $\phi=0.60$ studied in this work.
We observe that the first normal stress difference is small and still 
dominated by fluctuations.
Once more, the second normal stress difference is always negative and 
its absolute value exhibit a trend similar to the pressure and relative viscosity,
consistently to the ones observed in the literature \citep{MARI2014}.
The effect of the relative volume fractions of the two phases 
reflects the behaviour shown by the relative viscosity, with the exception of the first normal stress, similarly to what has been observed in \cref{fig:etaRgd}.

In order to highlight the origin of the reduced relative viscosity of the 
suspension with increasing dispersion ratio of the particles, we separate 
the contributions to the total stress accounted by the contacts, 
hydrodynamics and electrochemical potentials. In particular, \cref{fig:cont} 
shows the contributions to the relative viscosity for the suspensions 
at different dispersion ratios.
\Cref{fig:cont} is structured in a $4\times 3$ matrix of panels, 
where each row of the matrix shows the contributions to the relative
viscosity for increasing volume fractions ($\phi$ increases from top
to bottom), while each column shows the effect for different dispersion
ratios ($\lambda$ increases from left to right).
In each panel of the figure, we report the contribution to the
total relative viscosity (golden lines) of the contacts (black bars), 
hydrodynamics (darker grey) and electrochemical potentials (lighter grey), 
as a function of the shear-rate. \Cref{fig:cont} reveals that, at low shear rates 
(left end of each panel), the suspension rheology is dominated by the 
electrochemical potentials, with a little contribution due to the
hydrodynamics and a null contribution of the contacts. On the other hand, 
at high shear-rates (right end of each panel), the hydrodynamics has a 
small, almost shear-independent, contribution and the electrochemical 
potentials play no role, as expected; the contacts dominate the rheology 
of the suspension at these shear-rates, in agreement with the theory 
that relates shear-thickening and (frictional) contacts, recently proposed 
and experimentally proved 
\citep{SETO2013,FERNANDEZ2013,MARI2014,LIN2015,MARI2015}. 
\Cref{fig:cont} shows that, while the hydrodynamics remains constant in 
the shear-thickening regime and the consequent plateau, the contribution 
of the contacts drops when increasing the size of the larger phase 
dispersed. This phenomenon becomes more evident when dealing with larger 
volume fractions, where DST can be reduced to CST by simply adding 
larger particles (see the last row of panels of \cref{fig:cont}). 
This provides a quantitative explanation of the behaviour observed in 
the flow curves of \cref{fig:etaRgd}.

\begin{figure}[htb]
\centering%
\includegraphics[width=0.45\textwidth]{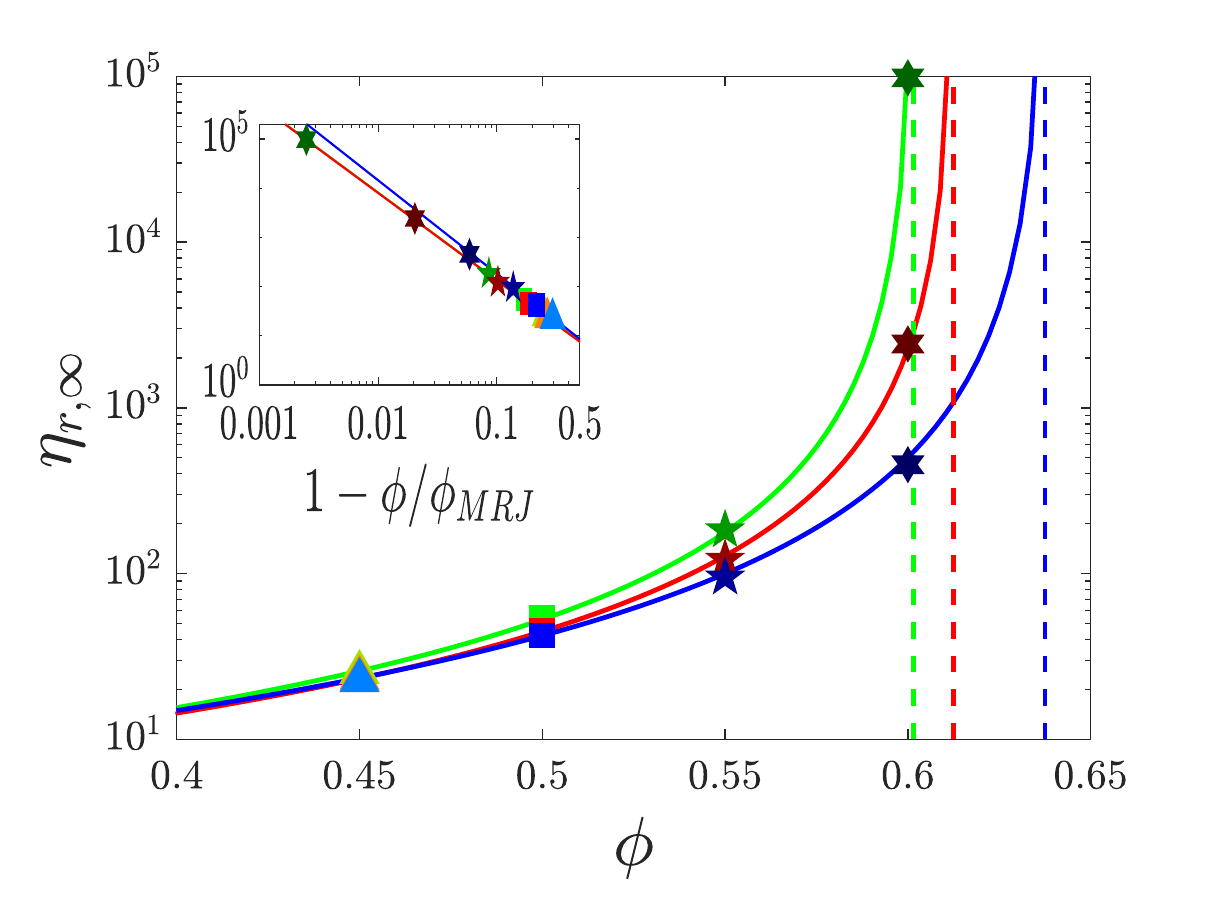}%
\caption{\label{fig:etaRphi} Relative viscosity in the high shear-rate limit, 
    $\eta_{r,\infty}=\eta_r(\dot{\gamma}/\dot{\gamma}_0 \to \infty)$, 
    as a function of the volume fraction $\phi$. The colour of the curves and 
    the style of the markers reflect the ones used in \cref{fig:etaRgd}. 
    The solid lines are obtained from fitting the data available with 
    power-laws $\eta_{r,\infty}=\alpha \left(1-\phi/\phi_{MRJ}\right)^{-\beta}$ 
    (as shown in the inset) and the dashed lines correspond to the values 
    of $\phi_{MRJ}$ obtained for the three configurations of suspensions considered. 
    The best fitting parameters are 
    $\left(\phi_{MRJ},\alpha,\beta\right)_{\lambda=1.5} =\left(0.6015,2.20,1.79\right)$,
    $\left(\phi_{MRJ},\alpha,\beta\right)_{\lambda=3.0} =\left(0.6125,2.18,1.79\right)$,
    $\left(\phi_{MRJ},\alpha,\beta\right)_{\lambda=6.0} =\left(0.6375,2.28,1.90\right)$.
    }%
\end{figure}
The reduction of contacts we observed for large values of $\lambda$, which is the 
origin of the lower relative viscosity, can be explained geometrically as a 
reduced total surface area of the suspension in the case with high dispersion 
ratios.  Indeed, as $\lambda$ grows, the total surface area 
$A=\pi a_1^2(n_1+n_2\lambda^2)$ monotonically decreases when the total volume 
fraction $\phi$ is kept constant (together with the relative volume fractions 
of the two phases that establish a non-linear dependency of $n_1$ and 
$n_2$ on $\lambda$).
In other terms, for a given volume fraction of the dispersed phases, the 
area available for contacts among particles is smaller for larger $\lambda$, 
thus reducing the probability of having contacts between the particles.
Note that this is valid only for spheres.
This result is confirmed by the coordination number defined as 
$N_c = 4 \pi \int_0^{\overline{r}/a_0}\! r^2\, g(r)\, N/V \, \mathrm{d}r$,
where $r$ is the distance from the origin, $V$ is the volume of the computational 
box, and $g(r)$ is the radial distribution function. 
Here, $\overline{r}/a_0=20$ is used to take into account the large dispersion ratio. 
The coordination number is an indicator of the average number of collisions between 
particles.
We computed $N_c$ for all the dispersion ratios of the suspensions at $\phi=60$ shown in 
\cref{fig:etaRphi} and, to emphasize the results, we normalised $N_c$ by the 
coordination number of the case with $\lambda=1.5$. Thus, we obtain $N_c=[1,0.859,0.856]$
for $\lambda=[1.5,3.0,6.0]$. As for the relative viscosity and the total surface area
$A$, the coordination number shows a decreasing trend with $\lambda$.
More quantitatively, the change of $A$ with $\lambda$ modifies the properties of 
the packing of the suspension as demonstrated by Hopkins \textit{et al.} 
\citep{HOPKINS2012,HOPKINS2013}. In particular, it was shown that the increase 
of the dispersion ratio $\lambda$ enhances the efficiency of the packing, 
resulting in an increase of $\phi_{MRJ}$. This geometrically constraint 
affects the rheological properties of the suspension when it approaches 
the jamming transition,  i.e.,\ when the contacts dominate the rheology. 
In this case, the relative viscosity $\eta_r$ can be expressed as a power 
law diverging for $\phi \rightarrow \phi_{MRJ}$, i.e.,\ $\eta_r \sim 
\left(1-\phi/ \phi_{MRJ} \right)^{-\beta}$ \citep{FORTERRE2008,BOYER2011,MARI2014}. 
We show this by considering the relative viscosity in the high shear-rate 
limit as a function of the volume fraction $\eta_{r,\infty}(\phi) = 
\eta_r(\phi,\dot{\gamma}/\dot{\gamma}_0\to \infty)$,  see \cref{fig:etaRphi}, 
where the plotted curves have been obtained by fitting our numerical data 
with a power-law of the form $\eta_{r,\infty}= \alpha \left(1-\phi/
\phi_{MRJ}\right)^{-\beta}$, as shown in the inset of the same figure 
\citep{MARI2014}. \Cref{fig:etaRphi} proves that $\phi_{MRJ}$ grows with 
$\lambda$, and thus explains that the differences in the values of the 
relative viscosity, previously observed at the beginning of this work 
for the different dispersion ratios (\cref{fig:etaRgd}), is a direct 
consequence of the variation of the packing efficiency which determines 
the infinite shear-rate limit of $\eta_r$ to be achieved after the 
shear-thickening region.

\section{Conclusion}
In this work, we studied the rheological behaviour of bidispersed 
suspensions of non-Brownian  spherical particles, and we showed that 
the presence of a large dispersed phase reduces the shear-thickening 
behaviour of the suspension.  This phenomenon is caused by the more 
efficient maximally random jammed packing exhibited by the suspensions 
with large dispersion ratio, that results in a reduction of contacts 
and thus of their contribution to the relative viscosity of the 
suspension. 
We also studied the effect of the relative volume fractions of the
dispersed phases and we found that a lower viscosity is achieved when
the dispersed phases have similar (identical in our work) volumes, 
coherently with the results already discussed in the literature.
The effect of the latter parameter on the rheological properties of the 
suspensions, however, is not as important as the bidispersion ratio.
The results discussed 
here are of particular interests as a simple control strategy for the 
rheology of particle suspensions,  especially in their shear-thickening 
regime, with direct impact on the flowability of the complex material. 
Indeed, it is possible to force the transition from DST to CST 
(or vice-versa) by simply substituting a partial volume fraction of a 
monodispersed suspension, with an equivalent one of large (or small) 
particles.

\backmatter

\bmhead{Acknowledgments}

All authors gratefully acknowledge the support of Okinawa Institute of Science and Technology Graduate University (OIST) with subsidy funding from the Cabinet Office, Government of Japan. The authors also acknowledge the computer time provided by the Scientific Computing section of Research Support Division at OIST.

\section*{Competing Interests}

The authors have no competing interests to disclose.


\end{document}